\documentclass{ws-ijmpa}

\usepackage[super]{cite}
\usepackage{xcolor}
\usepackage[verbose,hypertexnames=false]{hyperref}
\begin{document}

\markboth{Chengjun Xia, Xiaoyu Lai, and Renxin Xu}{Strange Matter}

%%%%%%%%%%%%%%%%%%%%% Publisher's Area please ignore %%%%%%%%%%%%%%%
%
\catchline{}{}{}{}{}
%
%%%%%%%%%%%%%%%%%%%%%%%%%%%%%%%%%%%%%%%%%%%%%%%%%%%%%%%%%%%%%%%%%%%%

\title{Strange Matter}

\author{Chengjun Xia}
\address{Center for Gravitation and Cosmology, College of Physical Science and Technology, Yangzhou University, Yangzhou 225009, P.R. China}

\author{Xiaoyu Lai}
\address{Institute of Astronomy and High Energy Physics, Hubei University of Education, Wuhan 430205, P.R. China}

\author{Renxin Xu}
\address{School of Physics and KIAA, Peking University, Beijing 100871, P. R. China\footnote{r.x.xu@pku.edu.cn}}

\maketitle

\begin{history}
\received{Day Month Year}
\revised{Day Month Year}
\accepted{Day Month Year}
\published{Day Month Year}
\end{history}

\begin{abstract}
Pulsar-like objects are extremely compact, with an average density that exceeds nuclear saturation density, where the fundamental strong interaction plays an essential role, particularly in the low-energy regime.
The internal structures and properties of those objects are profoundly connected to phenomena such as supernova explosions, gamma-ray bursts, fast radio bursts, high/low-mass compact stars, and even to issues like dark matter and cosmic rays.
However, due to the non-perturbative nature of quantum chromodynamics, significant uncertainties remain in our current understanding of the composition and equation of state (EOS) for the dense matter inside them.
Drawing on three-flavour symmetry and the strong coupling between light quarks, this paper presents a novel perspective on the nature of pulsars: they are actually composed of strange matter, in the form of either strange quark matter or strangeon (analogous to nucleons and representing multibaryon states with three-flavour symmetry) matter.
As both strange quark matter and strangeon matter contain non-zero strangeness, we refer to them collectively as ``strange matter'', and to the corresponding compact stars as ``strange stars''.
%
%Owing to the large mass of strangeons and the nucleon-like hard-core strong interactions between them, strangeon matter forms a crystalline structure at low temperatures, constituting a solid strangeon star.
We then briefly introduce several physical models describing strange matter and present the resulting structures and properties of strange stars. This includes discussions on the EOSs, surface properties, mass-radius relations, glitches, binary compact star mergers, and dark matter. Furthermore, we will explore how observational properties of pulsar-like objects support the strange star model.
\end{abstract}


\ccode{PACS numbers: 12.38.-t, 24.85.+p, 26.60.+c, 26.50.+x}

%12.38.-t Quantum chromodynamics
%24.85.+p Quarks, gluons, and QCD in nuclei and nuclear processes
%26.60.+c Nuclear matter aspects of neutron stars
%26.50.+x Nuclear physics aspects of novae, supernovae, and other explosive environments

\section{\label{sec:intro} Introduction}

You might not take strange things seriously, but is it really the case that strange matter is strange?
This is a question that is philosophically challenging to answer, as William Blake once said:\footnote{See the book ``$The~Marriage~of~Heaven~and~Hell$'' by William Blake (1757$-$1827), section ``Proverbs of Hell''.}
``What is now proved was once only imagined''.
We would argue that many independent thinkers throughout history held a similar view.
For example, the focus of this article is strongly relevant to the philosophy of $strangeness$: it was initially viewed with surprise, but today, following the successful establishment of the standard model of particle physics, it is considered just normal.

As one of the main components of the standard model of particle physics, quantum chromodynamics (QCD) is highly pertinent to the subject of this review.
Being one of the fundamental interactions in nature, QCD governs the strong interaction among quarks and gluons. In particular, nucleons are formed by confining three valence quarks within a bag via strong interaction, while the residue interaction plays a crucial role in binding nucleons together to form atomic nuclei. In this sense, ``nuclear/nucleon matter" should be termed ``strong matter" as its properties are dominated by strong interactions, whereas the ``strange matter" illustrated in this work also belongs to this category. Systems like nucleons and atomic nuclei are stabilized by strong interaction at the microscopic scale, while ``strange matter" is stabilized by strong interaction as well but at the macroscopic scale.
In contrast, normal matter composed of atoms can be termed ``electromagnetic matter" (or simply ``electric matter"), as electromagnetic interactions bind atomic nuclei and electrons into electrically neutral atoms and dominate their behavior. The long-range Coulomb repulsion between atomic nuclei prevents them from aggregating under normal pressure, rendering short-range strong interactions negligible in such matter.

Under extreme conditions, macroscopic ``strong matter" might form. In particular, during the late stages of stellar evolution, the immense gravitational force in the center overcomes the Coulomb repulsion, compressing electrons and nuclei into a dense state. The density of the dense state in the core exceeds those of atomic nuclei. What is the nature of the core at such a large density? From an observational perspective, these objects might appear as pulsar-like compact objects. Theoretically, they may be neutron stars, strange quark stars, or strangeon stars. The neutron star matter at their surfaces are comprised of ``electromagnetic matter", while both strange quark matter and strangeon matter are forms of ``strong matter" and exhibits greater stability.

In such cases, the study of compact star matter remains one of the most challenging frontiers in physics. In particular, understanding its physical nature has significant implications on various astrophysical phenomena such as supernovae, gamma-ray bursts (GRBs), fast radio bursts (FRBs), and even dark matter and cosmic ray detections. As will be illustrated in this paper, there is observational evidence for strange stars, but a definitive verification remains an open question.

This paper is organized as follows. We first review the history of neutron star and strange star research, then discuss their compositions and possible hybrid configurations. Models for their equations of state (EOS) are explored, focusing on aspects like surface properties, radii, glitches, and binary mergers. Observational tests supporting these models are summarized as well, followed by future prospects.

\section{\label{sec:neutron2strange} From Neutron Stars to Strange Stars}

The concept of compact stars dates back to Landau's 1932 proposal \cite{Landau1932_PZS1-285}, which suggested that stellar equilibrium could be maintained by converting electrons and protons into neutrons to form a dense core.\footnote{%
For a dense electron gas that follows the Fermi-Dirac statistics, an electron must populate a higher energy level if the lower levels are occupied, with a typical energy $\simeq \hbar c n^{1/3}\sim 10^2$~MeV if the electron number density $n\sim 1$~fm$^{-3}$.
This electron energy certainly disappears in neutron matter.
In this sense, Landau's neutron star is, in fact, a neutral star.
} %
However, Landau's model did not account for the equation of state of neutron matter, limiting its applicability.

In 1934, Baade and Zwicky hypothesized that supernovae resulted from the collapse of massive stars will form compact stars made of neutrons in the center, which will release significant amount of gravitational energy~\cite{Baade1934_PR46-76}. During stellar collapse, when the density of matter in the core regions exceeds the nuclear saturation density $\rho_0$ ($\simeq 2.8\times 10^{14}\rm\,g/cm^3$), the repulsive nuclear force and the degeneracy pressure of fermions will hinder the collapse. The sudden stoppage of infalling material generates a shockwave, ejecting outer layers into the space and leaving behind a compact star. This indicates the fact that if the core can resist further collapse without forming a black hole, a stable compact object with density above $\rho_0$ may form.

If protons~(p) and neutrons~(n) are elementary particles, then compact star matter with density above~$\rho_0$ must be neutron-rich. This is attributed to the large Fermi energy of electrons to attain charge neutrality, where protons will undergo inverse $\beta$-decay~$ e^-+p \rightarrow n+\nu_e$ to reduce the energy of the system. Numerical calculations have shown that the energy per baryon of neutron matter in vacuum is higher than that of iron, so neutron matter can not exist on the surface of the star, and there must be a transition region from neutron matter to normal matter inside a neutron star, i.e., the surface of the star is normal matter. At densities beyond $\sim2\rho_0$, exotic particles like pions, kaons, hyperons, and even quarks may emerge, challenging traditional neutron star models. Nevertheless, in all cases, these neutron star models always have a crust composed of normal matter bound by gravity.

Astronomers have discovered radio pulsars in the 1960s and soon realized that this was the neutron star predicted in theory~\cite{Hewish1968_Nature217-709}. In the meantime, particle physicists have found that hadrons such as protons and neutrons were not elementary particles, but were composed of quarks. Therefore, although astronomers at that time established the concept of ``pulsars are neutron stars" based on the experience of the previous 30 years, the possibility of quark degrees of freedom caused by the disappearance of hadronic structures in pulsars immediately attracted physicists' attention. In 1984, based on the study of the equilibrium and stability of stars composed of~u (upper), d (down) and s (strange) quarks \cite{Itoh1970_PTP44-291, Bodmer1971_PRD4-1601}, combined with the asymptotic freedom nature of the strong interaction between quarks, Witten proposed a conjecture: strange quark matter composed of nearly equal amounts of~u, d and s quarks may be more stable than iron~\cite{Witten1984_PRD30-272}.

\begin{figure}[t]
\centerline{\includegraphics[width=0.5\linewidth]{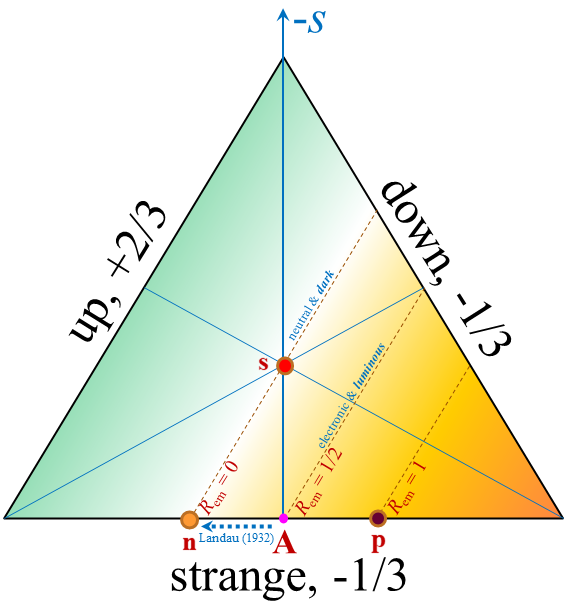}}
\caption{\label{Fig:triangle}The triangle of light-quark flavors. The points inside this triangle define the states with certain quark number densities of u, d and s quarks, indicated by the heights of one point to one of the triangle edges. The gray level denotes the charge-mass-ratio of quarks ($R$). Normal nuclei are around point A ($R=1/2$). Neutron stars are around point n and strange stars are around point s, both of which have nearly charge-neutrality ($R=0$). Another possible scenario is strangeon stars located also at point s, which are made of strangeons (see Section~\ref{sec:strange2strangeon}).}
\end{figure}

More specifically, the Fermi energy of the system can be reduced by increasing the degrees of freedom, where the system composed of~u, d, s quarks has lower energy~(i.e., more stable) than that of~u and d quarks alone. If the numbers of~u, d, s quarks are strictly equal, then the charge neutrality condition can be satisfied without electrons. In other words, a system with strictly equal numbers of~u, d and s can keep the Fermi energies of quarks and electrons as low as possible while maintaining flavor symmetry and charge neutrality\footnote{So why do ordinary nuclei have two flavors instead of three? Bigger is different! The reason is that the atomic nucleus in ordinary matter is microscopic, and the emergence of $s$-quark will increase the energy of the atomic nucleus~($m_s>m_ {u, d}$) and lead to its instability. Although strange sea quarks~($s\bar{s}$) may exist in the nucleus due to energy fluctuations, they cannot form strange valence quarks. Therefore, two-flavored atomic nuclei are energetically favored
as the electrons almost outside contribute negligible kinematic
energy.}. Fig.~\ref{Fig:triangle} illustrates the role of light flavor symmetry, where the fully symmetric point represented by point s corresponds to strange stars.

Strange stars may be formed in core collapse supernova, where baryonic matter in the core region of a massive star is being compressed drastically by gravity with the density reaching several times of the saturation density. The Fermi energy of electrons may make
%reach $\sim$300 MeV in this process, then it is
it energetically favorable to eliminate electrons via weak reactions, where two types of reactions may take place, i.e., the neutronization process ($e^- + p \to n + \nu_e$) and the strangenization process ($e^- + u \to s + \nu_e$ and $u + d \to s + u$).
%While neutronization eliminates electrons, strangenization can further reduce the energy and leads to the flavor symmetry for strong matter.
%
%At supra-nuclear densities, quark deconfinement may occur, with s quarks appearing via weak interactions ($u + d \leftrightarrow s + u$).
If strange quark matter is more stable than neutron matter and iron, the dense objects created in the supernova explosion may be strange quark stars composed of strange quark matter from the center to the surface. Due to QCD's complexity, Witten's conjecture remains unproven but motivates studies of strange quark matter.

Beside strange quark stars, strange stars could manifest as ``strangeon stars" (with quarks forming clusters). The MIT bag model~\cite{Alcock1986_ApJ310-261} describes quark confinement, while hybrid star models~\cite{Alford2005_ApJ629-969} incorporate both quark and nuclear phases. These configurations differ fundamentally in their internal interactions and surface properties. In conclusion, the debate on the nature of pulsars has formed two camps: neutron stars that can be seen as~``big nuclei'' and strange stars that can be seen as~``big hadrons". However, the strangeon stars mentioned below are different from these two types.

\section{\label{sec:strange2strangeon}From Strange Quark Matter to Strangeon Matter}

Strange matter composed of u, d, and s quarks may exist in various phases. Witten's ``strange quark matter" and Xu et al.'s ``strangeon matter" \cite{Xu2003_ApJ596-L59} represent two possibilities. The former involves a quark-gluon plasma, while the latter consists of quark clusters with residual strong interactions.

Using QCD-inspired models, Alford et al. explored color superconductivity in dense quark matter~\cite{Alford1998_PLB422-247}, while Fraga et al. incorporated perturbative QCD effects into bag models~\cite{Fraga2001_PRD63_121702}. However, achieving true deconfinement may require densities $\gtrsim 40\rho_0$, far beyond the typical density in neutron star cores. Therefore, the interactions between quarks in strange quark stars are expected to be sizable. If QCD is applied, the strong coupling constant~$\alpha_{\rm s}$ is too large to be used for perturbative expansion, so neither the color superconducting model nor MIT bag model is exactly self-consistent in theory.

On the one hand, it may be energetically favorable for strange quarks to exist in the interiors of pulsars. On the other hand, the interaction between quarks is still sizable even if the quarks inside the star exist in a deconfined phase of quark matter. In 2003, unlike quarks forming cooper pairs in momentum space under relatively~``weak'' strong interactions, Xu proposed that quarks in strange quark stars may aggregate into quark clusters under relatively~``strong'' strong interactions~\cite{Xu2003_ApJ596-L59}. In fact, baryons such as neutrons and protons can also be regarded as quark clusters, while the quark cluster inside the pulsar contains strange quarks and the baryon number may be greater than one. The quark cluster with strange quarks was later called~``strangeon" (i.e., nucleon-like unit with strange quarks).

In such cases, if we call matter with nearly equal numbers of~u, d, s quarks as strange matter, then there are two possible matter states: strange quark matter  and strangeon matter. The former is composed of three flavor free quarks, and the latter is composed of strangeons that confine the three flavor quarks. Both of them can be called~``strong matter". Correspondingly, stars composed of strange matter from the center to the surface are called strange stars. There are two types: strange quark stars and strangeon stars. Strangeon stars may be composed of strangeon matter from the center to the surface, or it may be that a strange quark matter core is wrapped in strangeon matter and forms a hybrid strangeon star~\cite{Zhang2023_PRD108-123031}.

Similar to the strange quark star, the strangeon star is a self-bound star, and its surface is a discontinuous interface separating the matter with density higher than~$\rho_0$ and the vacuum outside the surface. The self binding of strange quark stars comes from the strong interaction between quarks, while the self binding of strangeon stars comes from the strong interaction between strangeons. The former is similar to the interaction between quarks in nucleons, while the latter is similar to the interaction between nucleons. Unlike strange stars, neutron stars are gravitationally bound stars, which gradually turn from neutron matter into the outer crust material composed of normal matter as the density decreases from the center to the surface, and the density continuously drops to zero at the surface. Their distinct EOS and surface properties lead to different observational signatures.

%Strangeon stars with significant strangeness could form a solid state in compact stars.

\section{\label{sec:eos}Equation of State for Strange Matter}

The EOS for dense stellar matter is crucial in the study of compact stars and their internal structures. With negligible temperature effects, the EOS is often simplified into the relationship between pressure $P$ and energy density $\epsilon$, i.e., $P=P(\epsilon)$. To formulate the EOS, we need to know the basic constituents of matter and their interactions. Combined with the hydrostatic equations, we can then fix the macroscopic structures of compact stars such as their masses and radii. Since the nature of matter at supra-nuclear densities remains unknown, we often need to model the EOS and constrain it according to various observations. Various theoretical models were proposed to describe the properties of quark matter in the literature, e.g., perturbation model~\cite{Fraga2014_ApJ781-L25, Kurkela2014_ApJ789-127, Xu2015_PRD92-025025, Xia2017_NPB916-669, Xia2019_PRD99-103017}, linear sigma model~\cite{Holdom2018_PRL120-222001}, MIT bag model~\cite{Zhou2018_PRD97-083015, Miao2021_ApJ917-L22}, Dyson-Swinger equations~\cite{Roberts1994_PPNP33-477, Alkofer2001_PR353-281}, equivparticle model~\cite{Peng2008_PRC77-065807, Xia2014_PRD89-105027}, quasiparticle model~\cite{Pisarski1989_NPA498-423, Schertler1997_JPG23-2051, Schertler1997_NPA616-659}, and Nambu-Jona-Lasinio model~\cite{Buball2005_PR407-205, Gholami2025_PRD111-103034}. For strangeon matter, several EOS models have been proposed, including the polytropic model~\cite{Lai2009_AP31-128}, Lennard-Jones model~\cite{Lai2009_MMRAS398-L31}, corresponding state model~\cite{Guo2014_CPC38-055101},  Yukawa potential model~\cite{Lai2013_MNRAS431-3282}, and linked bag model~\cite{Miao2022_IJMPE0-2250037}.

\subsection{\label{subsec:MITbag}MIT bag model}
The strange quark matter has been examined in various QCD inspired models, where the MIT bag model is the most widely used one due to its simplicity. At fixed bag constant $B$, pairing gap  $\Delta$, strange quark mass $m_s$, and QCD correction parameter $a_4$, the thermodynamic potential density $\Omega$ of  quark matter can be written as~\cite{Fraga2001_PRD63_121702,Alford2005_ApJ629-969,Weissenborn2011_ApJ740-L14}:
\begin{equation}\begin{aligned}
\Omega=&-\frac{3a_4}{4\pi^2}\mu^4-\frac{\mu_{e}^4}{12 \pi^2}- \frac{12\Delta^2-3m_s^2}{4\pi^2}  \mu^2+B,
\label{eq:omega}
\end{aligned}\end{equation}
where $\mu_e$ is the electron chemical potential and $\mu=(\mu_u+\mu_d+\mu_s)/3$ the quark chemical potential with $\mu_u$, $\mu_d$ and $\mu_s$ being the chemical potentials for $u$, $d$ and $s$ quark, respectively. According to the basic thermodynamic relations, we can fix the pressure $P$, quark number density $n_{q}$, electron number density $n_{e}$, and energy density $\epsilon$, i.e.,
\begin{equation}
P=-\Omega, \,\, n_{q}=-\frac{\partial\Omega}{\partial \mu},\,\, n_{e}=-\frac{\partial\Omega}{\partial \mu_e},\,\,   \epsilon=\Omega+n_q \mu+n_e \mu_e ,
\label{eq:thermo}
\end{equation}
which gives:
\begin{eqnarray}
n_q&=&\frac{3a_4}{\pi^2}\mu^3 + \frac{2\lambda\sqrt{3a_4}}{\pi^2}\mu, \label{eq:rho_qe} \\ n_e&=&\frac{\mu_e^3}{3\pi^2}, \\
\epsilon&=&\frac{9a_4}{4\pi^2}\mu^4+\frac{\mu_{e}^4}{4 \pi^2}+B+ \frac{ \lambda\sqrt{3a_4} }{\pi^2}  \mu^2.
\label{eq:epson}
\end{eqnarray}
Finally, the EOS of quark matter reads~\cite{Pereira2018_ApJ860-12, Zhang2021_PRD103-063018}
\begin{equation}\label{eq:eos}
P=\frac{1}{3}(\epsilon-4B)+ \frac{4\lambda^2}{9\pi^2}\left(-1+{\mathrm{sign}(\lambda)}\sqrt{1+3\pi^2 \frac{(\epsilon-B)}{\lambda^2}}\right)
\end{equation}
with
\begin{equation}\label{eq:lam}
\lambda=\frac{c_1 \Delta^2-c_2 m_s^2}{\sqrt{c_3 a_4}},
\end{equation}
where the parameter set ($c_1$, $c_2$, $c_3$) take the values (1.86, 1, 0) for the 2SC phase; (3, 1, 0.75) for the 2SC+s phase; and (3, 3, 0.75) for the CFL phase, respectively~\cite{Zhang2021_PRD103-063018}.

\subsection{\label{subsec:polytropic}Polytropic Model}

The polytropic form $P=K\rho^{\Gamma}$ (where $K$ is a constant, $\Gamma=1+1/n$, and $n$ is the polytropic index), relating pressure $P$ to rest mass density $\rho$, finds wide application. It well describes white dwarfs and neutron stars and can also approximate non-relativistic degenerate matter. It is thus physically plausible describing the EOS of strange matter with a polytropic form~\cite{Lai2009_AP31-128}. In the framework of a polytropic model, the pressure and energy density are expressed as:
\begin{equation}
P=K\rho^{1+1/n}, \quad \epsilon=\rho c^2+nP.
\end{equation}
To account for the non-zero density at zero pressure (characteristic of self-bound matter like strange quark matter and strangeon matter), we introduce a vacuum energy density $\Lambda$ related to the QCD scale $\Lambda_{\rm QCD}$. The EOS can be rewritten as:
\begin{equation}
P=K\rho^{1+1/n}-\Lambda,
\label{eq:P_poly}
\end{equation}
\begin{equation}
\epsilon=\rho c^2+nK\rho^{1+1/n}+\Lambda,
\label{eq:e_poly}
\end{equation}
which ensures the density is non-zero at vanishing pressure. The MIT bag model can be seen as a special case with $n=3$, where $\Lambda$ corresponds to the bag constant $B$.

For a given polytropic index $n$, the vacuum energy density $\Lambda$,  the surface density $\rho_{\rm s}$, and the constant $K$ can be determined from the condition of zero pressure: $K=\Lambda \rho_{\rm s}^{-1-1/n}$. In practice, $\Lambda$ can be the same as the bag constant $B$ in MIT bag model, which is typically $\Lambda^{1/4} = 145$ MeV. The surface density $\rho_{\rm s}$ is usually set between 1.5 and 2 times $\rho_0=m_n n_0$ with $m_n$ the nucleon mass and $\rho_0$ the nuclear saturation density.

\subsection{\label{subsec:LJ}Lennard-Jones Model}

The polytropic model provides a phenomenological EOS without addressing the microscopic interactions among strangeons or quarks, where the Lennard-Jones potential introduced here offers a way to incorporate inter-particle interactions. We consider strangeons as multi-quark states, so that the interaction between color-singlet strangeons is primarily strong and short-ranged but share similar traits with the van der Waals forces between charge neutral atoms/molecules. Therefore, the Lennard-Jones potential is a suitable approximation for the interaction between two strangeons:
\begin{equation}
u(r)=u_0\left[\frac{4}{(r/r_0)^{12}}-\frac{4}{(r/r_0)^6}\right],
\label{eq:LJ}
\end{equation}
where $r$ is the distance between two strangeons, $u_0$ the potential depth, and $r_0$ the equilibrium distance. The first term represents the short-range repulsion, and the second term represents the long-range attraction. While the short-range repulsion might differ from the Pauli exclusion between quarks within hadrons, the Lennard-Jones form remains a useful approximation to account for the interaction between strangeons~\cite{Lai2009_MMRAS398-L31}.

Deriving the EOS for a many-body system interacting via pairwise potentials is complex. Calculating the total interaction energy requires summing over all pairs, often necessitating statistical approximations or simplifications. In nuclear physics, the mean-field approach approximates the interaction using an effective average potential, usually neglecting short-range correlations (e.g., the Woods-Saxon potential \cite{Ning2003}). For strangeon matter, we can assume each strangeon interacts only with its nearest neighbors. This is justified if the potential depth $u_0$ is sufficiently large, allowing strangeons to form a crystalline structure, i.e., a solid strangeon star and the crystal lattice naturally limits interactions to nearest neighbors.

Assuming the solid strangeon matter has a simple cubic (sc) lattice structure (common for close-packed solids; other structures yield qualitatively similar results), and given the pairwise potential Eq.~(\ref{eq:LJ}), the total potential energy for $N$ strangeons is then:
\begin{equation}
U(R)=2Nu_0\left[A_{12}\left(\frac{r_0}{r}\right)^{12}-A_6\left(\frac{r_0}{r}\right)^6\right],
\end{equation}
where $A_{12}=6.2$, $A_6=8.4$ \cite{Huang1988}, and $R$ is the distance between nearest neighbors. The number density $n$ relates to $R$ as $n=R^{-3}$. Substituting $r=R=1/n^{1/3}$ in the above equation gives $U$ in terms of $n$,  yielding the potential energy per particle:
\begin{equation}
\epsilon_{\rm p}=U/N=2u_0\left(A_{12}r_0^{12}n^5-A_6r_0^6n^3\right),
\label{eq:ep}
\end{equation}

For a large crystal, surface effects on the total potential energy and the mass-radius relation are negligible. The total energy density $\epsilon$ is the sum of the potential energy density and the rest mass energy density:
\begin{equation}
\epsilon=2u_0\left(A_{12}r_0^{12}n^5-A_6r_0^6n^3\right)+nm_{\rm c}c^2,
\label{eq:e_LJ}
\end{equation}
where $m_{\rm c}$ is the rest mass of a single strangeon. The pressure is derived according to thermodynamic relations, i.e.,
\begin{equation}
P=n^2\frac{{\rm d}(\epsilon/n)}{{\rm d}n}=4u_0\left(2A_{12}r_0^{12}n^5-A_6r_0^6n^3\right).
\label{eq:P_LJ}
\end{equation}

Now we fix the parameters in Eqs.~(\ref{eq:e_LJ}) and (\ref{eq:P_LJ}). Assuming each strangeon consists of $N_{\rm q}$ valence quarks, and each quark flavor (u, d, s) has a constituent mass roughly one-third of the nucleon mass $m_{\rm n}$, the mass per strangeon is then:
\begin{equation}
    m_{\rm c}=N_{\rm q}m_{\rm n}/3. \label{eq:mc}
\end{equation}
For color-singlet strangeons, $N_{\rm q}$ must be a multiple of 3. A typical configuration is the ``flavor-spin-color" symmetric state with $N_{\rm q}=18$, like the quark-$\alpha$ \cite{Michel1988_PRL60-677}. In fact, $N_q=18$ is favorable according to the state-of-art observations on the masses and radii of pulsars~\cite{Yuan2025_PRD111-063033}. The potential depth $u_0$ is expected to be on the order of $10^{2\text{-}3}$ MeV. The surface number density is obtained with $n_{\rm s}=[A_6/(2A_{12})]^{1/2}r_0^{-3}$ by setting pressure to zero. The baryon number density $n_{\rm b}$ is connected to the strangeon number density $n$ by $n_{\rm b}=n\cdot N_{\rm q}/3$, so that the surface baryon density is $n_{\rm bs}=N_{\rm q}[A_6/(2A_{12})]^{1/2}r_0^{-3}/3$. Then $r_0$ can be obtained for given $n_{\rm bs}$ (e.g., $n_{\rm bs}=2n_0$, where $n_0$ is nuclear saturation density). At given $N_{\rm q}$, $u_0$, and $n_{\rm bs}$, the EOS of strangeon matter can be fixed.

\subsection{\label{subsec:correspond}Corresponding State Model}

If the interaction between strangeons follows a Lennard-Jones potential as in Eq.~(\ref{eq:LJ}), the parameters $u_0$ and $r_0$ differ for different substances, leading to different EOS. However, by reducing pressure $P$, volume $V$, temperature $T$, and particle number $N$ using characteristic scales:
\begin{equation}
    P^*=Pr_0^3/u_0, \quad V^*=V/(Nr_0^3), \quad T^*=kT/u_0, \label{eq:duiying1}
\end{equation}
and introducing a quantum parameter $\Lambda^*$ reflecting quantum effects ($h$ is Planck's constant, $m$ is particle mass):
\begin{equation}
    \Lambda^*=h/(r_0\sqrt{mu_0}),      \label{eq:duiying2}
\end{equation}
then the reduced EOS $P^*$ vs. $V^*$ and $T^*$ for different substances obey a universal form:
\begin{equation}
    P^*=f(V^*, T^*, \Lambda^*),
\end{equation}
where $f$ contains substance independent constants. This is the principle of corresponding states, applicable to systems interacting via pairwise potentials. Using this principle, we can infer the EOS of an unknown system (like strangeon matter) from a known system (like noble gases) if they share the same functional form of interaction potential.

Since both strangeon matter and noble gases interact via Lennard-Jones potentials (as assumed in Sec.~\ref{subsec:LJ}), their reduced EOS are identical. We can thus fit the EOS of noble gases to experimental data and use the corresponding state principle to obtain the EOS for strangeon matter \cite{Guo2014_CPC38-055101}. This bypasses the need for detailed microscopic derivation.

Plotting $P^*$ vs. $\Lambda^*$ at fixed $V^*$ for different noble gases yields points lying approximately on a straight line. For strangeon matter, given $u_0$, $r_0$, and $N_{\rm q}$ (hence $m_{\rm c}$ via Eq.~(\ref{eq:mc})), we compute $\Lambda^*$ via Eq.~(\ref{eq:duiying2}). We then find $P^*$ from the noble gas line at that $\Lambda^*$ and $V^*$. Using Eq.~(\ref{eq:duiying1}), we obtain the number density $n=N/V$ and pressure $P$. Repeating for different $V^*$ values gives the $P-n$ relation. For example, with $u_0=40$ MeV, $r_0=2.5$ fm, $N_{\rm q}=18$, the $P-n$ relation is \cite{Guo2014_CPC38-055101}:
\begin{equation}
    P = (2.99\times 10^{41}n^{5.63} - 1.60\times 10^{34})\ \rm dyn/cm^2. \label{eq:P_duiying}
\end{equation}
Using $P=n^2{\rm d}(\epsilon/n)/{\rm d}n$, the energy density is:
\begin{equation}
    \epsilon=n\left(\int_{n_{\rm s}}^n\frac{P}{n^2} {\rm d} n+\frac{\epsilon_{\rm s}}{n_{\rm s}}\right) , \label{eq:e_duiying}
\end{equation}
where $\epsilon_{\rm s}$ is the surface energy density at surface number density $n_{\rm s}$. Combining this with Eq.~(\ref{eq:P_duiying}), the energy density $\epsilon(n)$ can be fixed.

At large temperatures, strangeon matter is expected to melt, transforming crystal structures into a liquid phase. The melting heat per particle $H$ with respect to the quantum parameter $\Lambda^*$ can also be fixed using the corresponding state model, i.e.,
\begin{equation}
    H=1.18 u_0 \exp{\left[-\left(\frac{\Lambda^*-0.12}{1.60}\right)^2\right]}, \label{eq:melt}
\end{equation}
which is obtained by fitting to those of ordinary substances such as Xe, Kr, Ar, Ne, H$_2$, He~\cite{Guo2014_CPC38-055101}. This gives typically $H\sim$ 10 MeV for the strangeon matter considered here, which could be responsible for the plateau observed in various $\gamma$-ray bursts~\cite{Dai2011_SCPMA54-1541}.

\subsection{\label{subsec:H}Yukawa Model}

What constitutes strangeon matter? While the $\Lambda$ hyperon (uds) fulfills the flavor symmetry of QCD and is unstable, the H-dibaryon (uuddss), a bound state of two $\Lambda$s, may be stable \cite{Jaffe1977_PRL38-195}. Lattice QCD simulations suggest that the H-dibaryon might be stable with a binding energy of tens of MeV \cite{Beane2011_PRL106-162001, Inoue2011_PRL106-162002}. This encourage us to consider stars composed of H-dibaryons~\cite{Lai2013_MNRAS431-3282}.

Taking H-dibaryons as constituents, we model the interaction between two H-dibaryons separated by distance $r$ using Yukawa potentials mediated by $\sigma$ and $\omega$ mesons~\cite{Faessler1997_PLB391-255}, i.e.,
\begin{equation}
    V(r)=\frac{g_{\omega H}^2}{4\pi}\frac{e^{-m_\omega r}}{r}-\frac{g_{\sigma H}^2}{4\pi}\frac{e^{-m_\sigma r}}{r}, \label{V_H}
\end{equation}
where $m_\omega$ and $m_\sigma$ are the masses of the $\omega$ and $\sigma$ mesons, and $g_{\omega H}$, $g_{\sigma H}$ the corresponding coupling constants between the mesons and H-dibaryon. Note that the precise values for $g_{\omega H}$ and $g_{\sigma H}$ are unknown, while here we assume they scale with the nucleon couplings $g_{\omega N}$ and $g_{\sigma N}$.

For dense matter composed of H-dibaryons, medium effects become important. We incorporate the Brown-Rho scaling \cite{Brown1990_PLB237-3, Brown2004_PR396-1}, where meson and baryon masses decrease with density:
\begin{equation}
    \frac{m^*_\omega}{m_\omega}=\frac{m^*_\sigma}{m_\sigma}=\frac{m^*_H}{m_H}=1-\alpha_{\rm BR}\frac{n}{n_0},
\end{equation}
Here $m^*_\omega$, $m^*_{\sigma}$, $m^*_H$ are the effective masses of the $\omega$ meson, $\sigma$ meson, and H-dibaryon, $n_0$ is nuclear saturation density, $n$ is the H-dibaryon number density, and $\alpha_{\rm BR}$ is a scaling parameter. With Brown-Rho scaling and suitable parameters, H-dibaryon stars may be more stable than neutron stars~\cite{Lai2013_MNRAS431-3282}.

Similar to Sec.~\ref{subsec:LJ}, the short-range repulsion in the potential favors crystallization. We make analogous assumptions: the matter forms a sc lattice, and only nearest-neighbor interactions are significant. The potential energy per particle is then:
\begin{equation}
    \epsilon_{\rm p}=\frac{1}{2}n\left(A_{12}\frac{g_{\omega H}^2}{4\pi}\frac{e^{-m^*_\omega R}}{R}-A_6\frac{g_{\sigma H}^2}{4\pi}\frac{e^{-m^*_\sigma R}}{R}\right),
\end{equation}
where $A_{12}$ and $A_6$ take values as in Sec.~\ref{subsec:LJ}, and $R$ is the nearest-neighbor distance ($R=n^{-1/3}$). Applying the lattice sum gives the total energy density:
\begin{equation}
    \epsilon=\frac{1}{2}n^{4/3}\left(A_{12}\frac{g_{\omega H}^2}{4\pi}e^{-m^*_\omega n^{-1/3}}-A_6\frac{g_{\sigma H}^2}{4\pi}e^{-m^*_\sigma n^{-1/3}}\right)+nm^*_H,
\end{equation}
In this work, the H-dibaryon mass in vacuum is taken as $m_H=2210$ MeV, with $\alpha_{\rm BR}$ lies between 0.1 and 0.2 \cite{Brown1991_PRC44-2653}. Finally, the pressure can be fixed by
\begin{equation}
   P=n^2{\rm d}(\epsilon/n)/{\rm d}n.
\end{equation}

\subsection{\label{subsec:linkedbag}Linked Bag Model}

\begin{figure}[t]
\centerline{\includegraphics[width=0.5\linewidth]{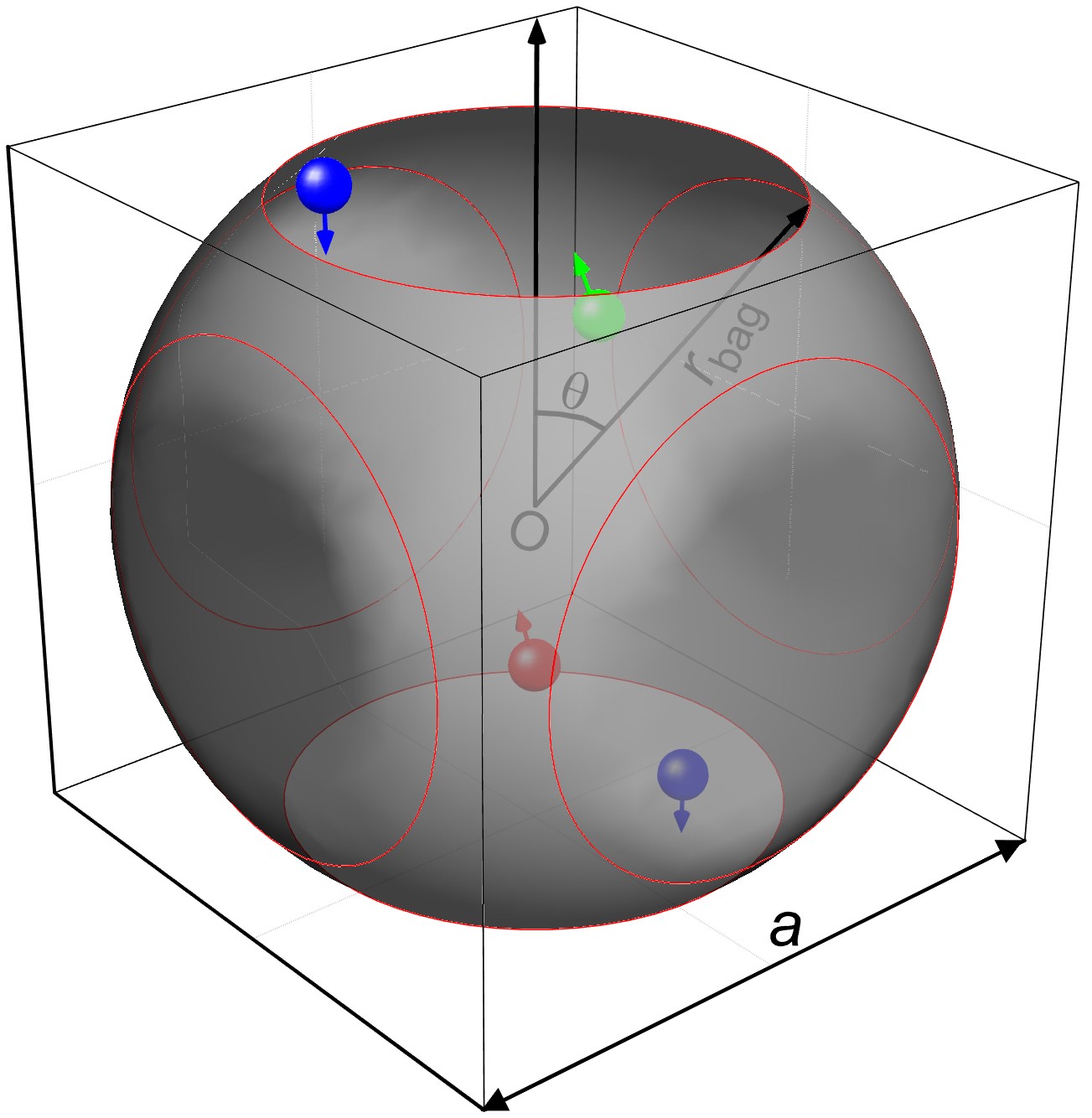}}
\caption{\label{fig:bag_lattice} A schematic illustration of a lattice cell in strong matter. A spherical bag (centered at point ``O'') resides in the cell, which is linked to the neighbouring bags through the six windows (the red circles) on the bag's surface. The size of each window is characterized by the angle $\theta$ with $\cos\theta = a/(2r_{\rm bag})$, where $a$ is the lattice constant and $r_{\rm bag}$ the bag radius. Taken from Ref.~\citen{Miao2022_IJMPE0-2250037}.}
\end{figure}

In the linked bag model, strangeon matter consists of bags, each containing $N_{\rm q}$ valence quarks, confined within a radius $r_{\rm bag}$. The bags are arranged in a crystalline lattice with lattice constant $a$. The baryon number density $n_{\rm b}$ relates to $a$ via $a=(A/n_{\rm b})^{1/3}$, where $A=N_{\rm q}/3$ is the baryon number per bag. Fig.~\ref{fig:bag_lattice} illustrates the geometry. When $r_{\rm bag}>a/2$, adjacent bags overlap, deforming their spherical shapes. The overlapping region forms a ``neck" with an opening angle $\theta=\arccos(a/2r_{\rm bag})$. For $r_{\rm bag}\geq \sqrt{3}a/2$ ($\theta\geq54.7^\circ$), the neck vanishes, suggesting a possible phase transition to quark matter.

Using the multiple reflection expansion (MRE) approximation~\cite{Berger1987_PRC35-213, Madsen1993_PRL70-391, Madsen1993_PRD47-5156, Madsen1994_PRD50-3328}, the total energy per unit cell is:
\begin{equation}
\label{eq:energy per cell}
E=\sum_{j}(\Omega_j+N_j\mu_j)+BV-\frac{z_0}{r_{\rm bag}}\frac{\omega}{4\pi},
\end{equation}
where $\Omega_j$, $N_j$, $\mu_j$ are the thermodynamic potential, particle number, and chemical potential for particle species $j$, respectively. $B$ is the bag constant, $V$ is the enclosed volume of the bag, and the last term represents the zero-point energy with $z_0$ being a phenomenological constant. The geometric factor $\omega$ depends on $r_{\rm bag}$ and $a$:
\begin{equation}
  \omega =
 \left\{\begin{array}{l}
   4\pi,\  r_{\rm bag}<\frac{a}{2}\\
   4\pi\left(\frac{3a}{2r_{\rm bag}}-2\right),\  \frac{a}{2}\leq r_{\rm bag}<\frac{\sqrt{2}a}{2}\\
   \int_{\theta_1}^{\theta_2} \cos \theta \left[\frac{\pi}{2}-2\cos^{-1}\left(\frac{a}{2r_{\rm bag}\cos \theta}\right)\right]\mbox{d}\theta,\  \frac{\sqrt{2}a}{2}\leq r_{\rm bag}<\frac{\sqrt{3}a}{2}\\
   0,\  r_{\rm bag}\geq\frac{\sqrt{3}a}{2}\\
 \end{array}\right.. \label{Eq:omega}
\end{equation}
The MRE formalism decomposes the thermodynamic potential $\Omega_i$ for species $i$ into volume ($\Omega_{i,V}$), surface ($\Omega_{i,S}$), and curvature ($\Omega_{i,C}$) terms:
\begin{eqnarray}
\Omega_{i,V}&=& -\frac{g_i V}{24\pi^2}\left[\mu_i u_i(\mu_i^2-\frac52m_i^2)+\frac32m_i^4\ln\frac{\mu_i+u_i}{m_i}\right]
            +\frac{g_i\alpha_s}{12\pi^3}\left[ -2u_i^4 \right. \label{eq:Omega_V}\\
            &&{}\left. + 3\left(\mu_i m_i-m_i^2\ln\frac{\mu_i+u_i}{m_i}\right)^2 +\left(6m_i^2\ln\frac{\bar\Lambda}{m_i}+4m_i^2\right)\left(\mu_i u_i - m_i^2 \ln\frac{\mu_i+u_i}{m_i}\right)\right], \nonumber \\
\Omega_{i,S}&=&\frac{-g_i S}{8\pi}\left[\frac{1}{3\pi}\left(\mu_i^3\arctan\frac{u_i}{m_i}-2\mu_i u_i m_i  + m_i^3\ln\frac{\mu_i+u_i}{m_i}\right) -\frac{\mu_i u_i^2}{6} \right.\nonumber\\
            &&{}\left.+\frac{m_i^2(\mu_i-m_i)}{3} \right], \label{eq:Omega_S} \\
\Omega_{i,C}&=&\frac{g_i C}{48\pi^2}\left(m_i^2\ln\frac{\mu_i+u_i}{m_i}+\frac{\pi}{2}\frac{\mu_i^3}{m_i}-\frac{3\pi\mu_im_i}{2}+\pi m_i^2  -\frac{\mu_i^3}{m_i}\arctan\frac{u_i}{m_i}\right), \label{eq:Omega_C}
\end{eqnarray}
where $u_i\equiv\sqrt{\mu_i^2-m_i^2}$, $g_i$ is the degeneracy (e.g., $g_u=g_d=g_s=6$ for quarks). The surface area $S=\omega r_{\rm bag}^2$ and curvature $C=2\omega r_{\rm bag}$ define the geometric factors.

The bag constant $B$ encodes non-perturbative QCD effects. While perturbative QCD suggests $B\simeq455\,{\rm MeV/fm^3}$~\cite{Shuryak1978_PLB79-135}, phenomenological fits according to baryon spectra often indicates much smaller values like $B\simeq 50\,{\rm MeV/fm^3}$~\cite{DeGrand1975_PRD12-2060}. In such cases, we take $B$ dependent on the chemical potentials:
\begin{equation}
B=B_0+B_2\xi^2+B_3\xi^3,
\end{equation}
with  $\xi=(\sum_i N_i\mu_i/A-m_N)/m_N$, $B_0=50\,{\rm MeV/fm^3}$, and $B_2$, $B_3$ as parameters. The particle numbers $N_j$ relate to the chemical potentials $\mu_j$ via:
\begin{equation}
\label{eq:Ni}
N_j=-\frac{\partial \Omega_j}{\partial \mu_j}-\frac{\partial B}{\partial \mu_j}V.
\end{equation}

For a fixed lattice constant $a$, the bag radius $r_{\rm bag}$ and particle numbers $N_i$ are determined by minimizing the total energy $E$ per cell. The energy per baryon is $E/A$, which determines the energy density:
\begin{equation}
\varepsilon=n_\mathrm{b}E/A. \label{eq:energy density}
\end{equation}
Finally, the baryon chemical potential and pressure are derived according to basic thermodynamic relations, i.e.,
\begin{eqnarray}
\mu_{\rm b}&=&\frac{d \varepsilon}{d n_\mathrm{b}}, \\
P&=&n_\mathrm{b}^2\frac{d}{d n_\mathrm{b}}\frac{\varepsilon}{n_\mathrm{b}}=n_\mathrm{b} \mu_{\rm b}-\varepsilon.
\label{eq:eos}
\end{eqnarray}

\section{How to test the strange star model?}

Over the past several decades, utilizing telescopes like Parkes (64 m), Arecibo (305 m), and China's FAST (500 m), radio pulsar surveys have yielded rich observational data. Over 4000 pulsars are currently known, most classified as rotation-powered pulsars. Others pulsar-like objects include X/$\gamma$-ray pulsars, and objects with no clear pulsed radio emission, like Anomalous X-ray Pulsars (AXPs), Soft Gamma-ray Repeaters (SGRs), X-ray Dim Isolated Neutron Stars (XDINS), and Central Compact Objects (CCOs). Different internal structure models predict distinct observational signatures, where in this section we will examine evidence for strangeon stars from surface properties, mass-radius relations, glitches, and binary compact star mergers.

\subsection{Surface Properties}

The surface of a strange star is dominated by strong interactions, potentially leading to phenomena distinct from neutron stars. A key example is the origin of thermal X-ray emission from isolated pulsars. For neutron stars, surface layers composed of light elements (H/He) are expected to produce observational thermal X-rays. Missions like Chandra and XMM-Newton have searched for spectral lines from these elements but found none. This is naturally explained if pulsars' surfaces lack atomic structures, consistent with the solid strangeon star model or strange quark star model.

The strong surface magnetic field of pulsars powers their magnetospheric activity. The Ruderman-Sutherland (RS) model is widely used, describing gap discharge above polar caps. However, in neutron stars, the surface binding energy for ions might be insufficient for such discharge unless in the presence of super-strong magnetic field ($B_q=4.14\times10^{13}$ G). For strangeon stars, the charge separation mechanism differs significantly \cite{Xu1999_ApJ522-L109, Yu2010_RAA10-815}, and could potentially explain drifting subpulses~\cite{Lu2019_SCPMA62-959505}. Furthermore, the solid surface allows for ``mountain"-like structures triggering localized discharges, whose dynamics can reveal their surface properties~\cite{Xu2025}. In fact, general pulse profiles non-symmetrical to the meridian plane on which the rotational and magnetic axes lie could result from preferential sparking around small mountains on stellar surfaces~\cite{Xu2024_AN345-e230153}.

Strong binding also impacts surface emissivity: pulsar emission isn't limited by the Eddington limit, allowing super-Eddington bursts like those in Gamma-Ray Bursts (GRBs)~\cite{Chen2007_ApJ668-L55}. Due to the presence of a strangeness barrier, the ``atmosphere" of a strange star could be regarded as the upper layer of a normal neutron star and can produce characteristic X-ray ``bands" and optical ``bands" in spectra observed in XDINS~\cite{Wang2017_ApJ837-81}, the O VIII Ly-$\alpha$ line, and the black-body emission of the X-ray binary 4U 1700+24~\cite{Xu2014_RAA14-617}. Additionally, absorption features at 0.7 keV and 1.4 keV in CCO 1E 1207.4-5209 may correspond to oscillation frequencies of electrons bound to the strange star's surface~\cite{Xu2012_PRD85-023008}.

\subsection{Mass and Radius of Compact Stars}

The hydrostatic equilibrium equation in general relativity (Tolman-Oppenheimer-Volkoff equation, TOV) is given by:
\begin{equation}
\frac{{\rm d}P}{{\rm d}r}=-\frac{Gm(r)\epsilon}{r^2c^2}\frac{\left(1+\frac{P}{\epsilon}\right)\left(1+\frac{4\pi r^3P}{m(r)c^2}\right)}{1-\frac{2Gm(r)}{rc^2}},
\label{eq:TOV}
\end{equation}
where $m(r)=\int_0^r 4\pi r^2 \epsilon {\rm d} r$ is the mass enclosed within radius $r$. To fix the stellar structure, we integrate the TOV equation outward from a central density $\rho_{\rm c}$, using the appropriate EOS at each density, until pressure drops to zero at the surface. This yields the total mass $M$ and radius $R$. Varying $\rho_{\rm c}$ generates a curve on the $M$-$R$ plane, and a curve on the $M$-$R$ diagram corresponds to a specific EOS model.

As central density $\rho_{\rm c}$ increases, the mass of compact stars generally increase until reaching a maximum value $M_{\rm max}$. Further increasing $\rho_{\rm c}$ leads to unstable configurations where a slight perturbation will cause the star oscillates exponentially. This maximum stable mass is called TOV mass, i.e., $M_{\rm TOV}= M_{\rm max}$. Different EOS models yield different $M$-$R$ curves and $M_{\rm TOV}$. If $M_{\rm TOV}$ predicted by a model is less than the observational masses of massive pulsars, then the model can be ruled out.

\begin{figure}[t]
\centerline{\includegraphics[width=0.7\linewidth]{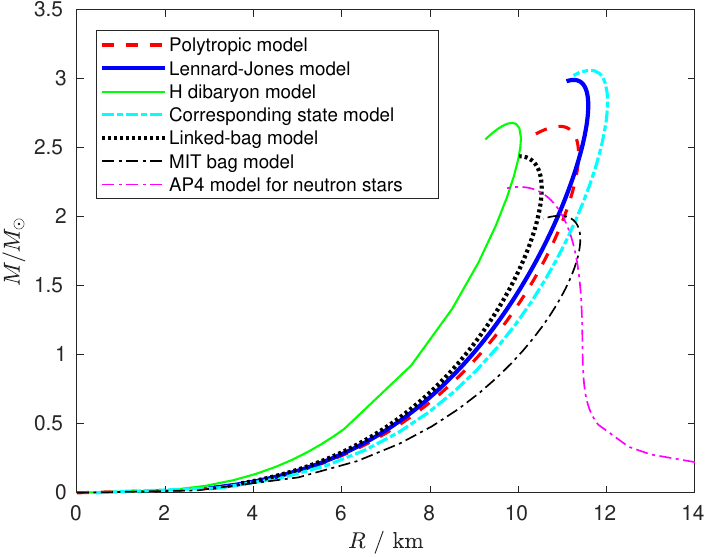}}
\caption{\label{fig:MR} The~$M$-$R$ curves of strange stars predicted by various models. The black dot-dashed line shows the results of the MIT bag model ($B^{1/4}=145\ \rm MeV$, $\Delta=0$, $a_4=1$), the red dashed line represents the polytropic model~($n=0.5$), the blue solid line represents the Lennard-Jones model~($n_{\rm s}=0.36\ \rm /fm^3$, $u_0=30$ MeV, $N_{\rm q}=18$), the green solid line represents the H dibaryon model~($\alpha_{\rm BR}=0.15$, $n_{\rm s}=2n_0$), the cyan dash-dotted line represents the corresponding state model~($u_0=40$ MeV, $r_0=2.5$ fm, $N_{\rm q}=18$), and the black dotted line represents the linked-bag model~($B_2=162.3\ \rm MeV/fm^3$, $B_3=100\ \rm MeV/fm^3$, $z_0=2.843$). For comparison, the pink dot-dashed line shows the results of neutron star model~AP4.}
\end{figure}

Figure~\ref{fig:MR} shows the $M$-$R$ curves predicted by the EOS models listed in Sec.~\ref{sec:eos}. The black dot-dashed line indicates strange stars predicted by the MIT bag model  ($B^{1/4}=145\ \rm MeV$, $\Delta=0$, $a_4=1$). The red dashed line represents a polytropic model with $n=0.5$. The blue solid line represents a Lennard-Jones model with $n_{\rm s}=0.36\ \rm fm^{-3}$, $u_0=30$ MeV, $N_{\rm q}=18$. The green solid line represents an H-dibaryon model with $\alpha_{\rm BR}=0.15$, $n_{\rm s}=2n_0$. The cyan dash-dotted line represents a corresponding state model with $u_0=40$ MeV, $r_0=2.5$ fm, $N_{\rm q}=18$. The black dotted line represents a linked bag model with $B_2=162.3\ \rm MeV/fm^3$, $B_3=100\ \rm MeV/fm^3$, $z_0=2.843$. For reasonable parameters, models like Lennard-Jones, H-dibaryon, and corresponding state can support $M_{\rm TOV} > 3 M_\odot$. For comparison, results for neutron stars predicted by the AP4 model~\cite{Akmal1997_PRC56-2261} are shown as pink dot-dashed line, whose TOV masses are smaller but still exceed the two-solar-mass constraints~\cite{Antoniadis2013_Science340-1233232}. Note that masses in the figure are for non-rotating stars, while rotation increases the maximum mass $M_{\rm max}$ above $M_{\rm TOV}$~\cite{Yang2024_RAA24-035005}.

The gravitational wave event GW170817 from a binary compact star merger (see Sec.~\ref{subsec:merge}) has provided detailed information on tidal deformation, constraining the radii of compact stars at intermediate masses ($\sim 1.4 M_\odot$). This offers a possibility to distinguish between the EOS models: models predicting smaller radii are favored to explain the binary neutron star merger GW170817, and also support a sufficiently high $M_{\rm TOV}> 2 M_\odot$. While the GW170817 signal and the existence of $2 M_\odot$ pulsars rule out some soft EOS (like AP4), the possibility of massive ($\sim 2.5 M_\odot$) components in future merger events remains open, supporting stiff EOS models~\cite{Annala2018_PRL120-172703}.

While searching for massive pulsars, finding low-mass ($<1M_\odot$) compact stars also tests EOS models. For neutron stars, the minimum mass is $\sim 0.1 M_\odot$ \cite{Haensel2002_AA385-301}, too low to reach supra-nuclear densities. For strangeon stars, the minimum mass could be much smaller. The 2022 mass measurement of the CCO candidate HESS J1731-347 / XMMU J173203.3-344518 yielded $M\simeq 0.77 M_\odot$, $R\simeq 10.4$ km~\cite{Doroshenko2022_NA6-1444}, while an earlier measurement of 4U 1746-37 also indicates an ultra low mass and small radius compact object~\cite{Li2015_ApJ798-56}. Such a small radius for a low mass compact star challenges neutron star models but is natural for strange stars.

\subsection{Glitches and Starquakes}

As a strangeon star cools down with its temperature drops below a critical value (around $10^{9-10}$ K for crystallization), the entire star solidifies~\cite{Xu2003_ApJ596-L59}. In a rotating solid star, crustal strain will eventually build up due to spin-down. When stress reaches a critical threshold, a sudden release of strain energy takes place, which leads to starquakes. Meanwhile, the solid core may also store significant strain energy, whose sudden release could power extreme events like GRBs~\cite{Xu2006_MNRAS373-L85, Xu2009_SCPMA52-315, Chen2024_RAA24-025005} or repeating FRBs \cite{2022SCPMA..6589511W, 2022ApJ...927..105W, 2022MNRAS.517.5080W, Xu2024_AN345-e230153}.

Glitches can also originate from starquakes, which are crucial to probe the internal structures of pulsars. Due to magnetic braking, a pulsar's spin frequency $\nu$ decreases gradually. Occasionally, $\nu$ suddenly increases by $\Delta\nu$ (within minutes) and then recovers (over days/weeks). This phenomenon is called a glitch, where the observed glitch sizes $\Delta\nu/\nu$ range from $10^{-10}$ to $10^{-5}$.

The unknown internal structures of pulsars lead to various possible glitch mechanisms. A leading model attributes the sudden spin-up to angular momentum transfer from a superfluid component. However, crustal superfluid models struggle to explain large glitches like those observed in Vela pulsar ($\Delta\nu/\nu \sim 10^{-6}$). Even considering angular momentum transfer between the superfluid and normal parts throughout the star, the available angular momentum from crustal superfluid vortices is insufficient for large glitches~\cite{Andersson2012_PRL109-241103}.

\begin{figure}[t]
\centerline{\includegraphics[width=0.7\linewidth]{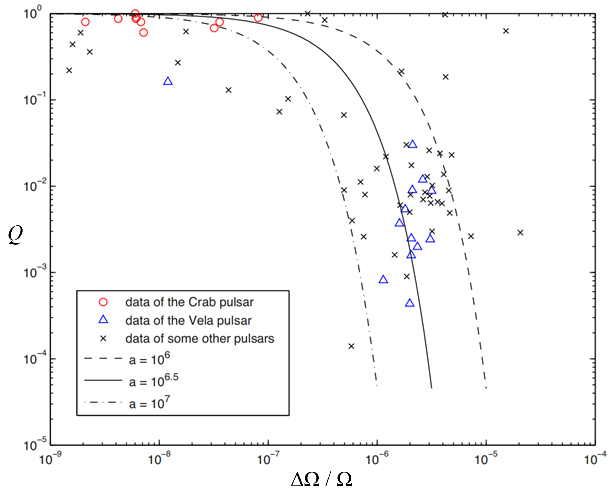}}
\caption{\label{fig:Q} The recovery coefficient~$Q$ as functions of glitch size~$\Delta \Omega/\Omega$, where the parameter~$a$ represents different degrees of exponential recovery. The observational values for the Crab pulsar, the Vela pulsar and several other pulsars are indicated by the red circles, blue triangles and black crosses, respectively. Taken from Ref.~\citen{Lai2018_MNRAS476-3303}.}
\end{figure}

For solid strangeon stars, starquakes releasing elastic energy become a plausible glitch mechanism. Models relating glitch size and waiting time (time between glitches) to the shear modulus and critical stress of solid strangeon matter have been proposed~\cite{Zhou2004_AP22-73}, where the glitches in strangeon stars~\cite{Peng2008_MNRAS384-1034} and energy release processes~\cite{Zhou2014_MNRAS443-2705} were extensively investigated. By dividing the inner motion of the star into plastic flow and elastic motion, as indicated in Fig.~\ref{fig:Q}, the glitch sizes and recovery behaviors observed in Crab and Vela pulsars can be understood in a unified model despite the 3 orders of magnitude differences in their glitch sizes~\cite{Lai2018_MNRAS476-3303}.

Similar to earthquake models, our recent work incorporates the effect of internal stress build up in rotational dynamics, where the instantaneous change in spin period for quakes take place at various depths were analysed~\cite{Lu2023_MNRAS520-4289}. It was found that while the trigger mechanism for starquakes might involve rotational dynamics, the solid core itself provides the necessary stress reservoir. The much higher shear modulus in the strangeon core compared to the neutron star crust allows for larger stress drops, enabling large glitches up to $\sim 10^{-6}$.

\begin{figure}[t]
\centerline{\includegraphics[width=0.7\linewidth]{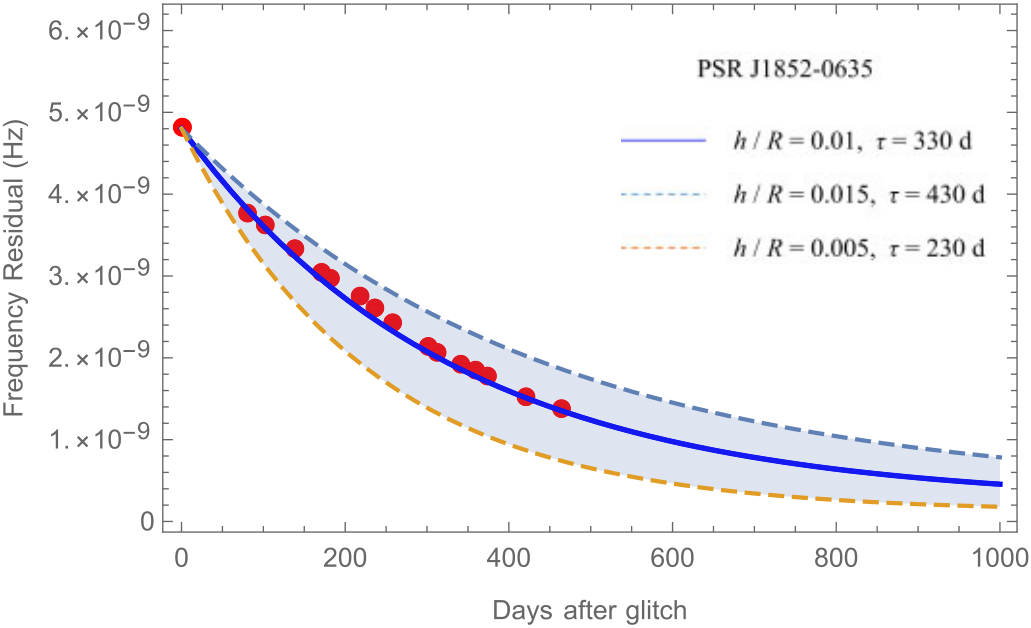}}
\caption{\label{fig:recovery} Frequency residual in the recovery for a glitch of PSR J1852-0635 in the framework of strangeon star model, where $h$ is the depth of the cracking site below the surface, $R$ the stellar radius, and $\tau$ the viscous time-scale. The red solid circles are the data points extracted from observation, while the first circle at $t=0$ is derived from the glitch size and before the glitch. Taken from Ref.~\citen{Lai2023_MNRAS523-3967}.}
\end{figure}

For the post-glitch recovery behavior, a model involving viscous flow in fractures has been proposed~\cite{Lai2023_MNRAS523-3967}. After the quake, the pressure at the slip fracture drops significantly, and residual pressure gradients drive fluid motion within the fracture. The plastic motion of the outer material and the elastic motion of the inner material during a starquake may break some of the strangeon matter into small pieces, which will naturally fill the places with low pressure. Since the fragmentation during starquake occurs at the equatorial plane, this ``fluid" flow will increase strangeon star's moments of inertia and reduce the rotation frequency, corresponding to the recovery behavior after the glitch. As indicated in Fig.~\ref{fig:recovery}, the movement of fragments can be described as viscous fluid, and it was shown that the change of rotation frequency with time is very close to the exponential recovery form, which is consistent with the observational glitch recovery.

\subsection{\label{subsec:merge}Mergers of Binary Strange Stars}

LIGO's first gravitational wave detection, GW150914, marked the era of gravitational wave astronomy. The binary neutron star merger GW170817 \cite{LVC2017_PRL119-161101} provides a new avenue to probe a compact star's internal structure. %This is a landmark discovery in multi-messenger astronomy.

\subsubsection{\label{subsubsec:tidal}Inspiral: tidal deformability}

As a binary system inspirals, tidal deformation becomes significant in the late stages.
%For strangeon stars, the strong inter-strangeon binding makes them less deformable than neutron stars of the same mass.
The deformability is quantified by the dimensionless tidal deformability $\Lambda=(2/3)k_2/(GM/c^2R)^5$, where $M$ and $R$ are the star's mass and radius, and $k_2$ is the second tidal Love number and dependent on the EOS. A smaller $\Lambda$ means that the star is stiffer and less deformable under a given tidal field, while for a larger $\Lambda$ the star is more deformable.
Because tidal deformation alters the orbital motion compared to point-mass inspiral and hence encode information about $\Lambda$ in the waveforms of gravitational waves, it can be used to constrain EOS directly. In particular, strange stars are significantly stiffer than neutron stars. The density drops rapidly from the core to the surface in neutron stars, while the strong binding maintains large density even near the surface in strange stars.
Therefore, a strange star is smaller and has a smaller $\Lambda$ compared with a neutron star of same mass.

\begin{figure}[t]
\centerline{\includegraphics[width=0.7\linewidth]{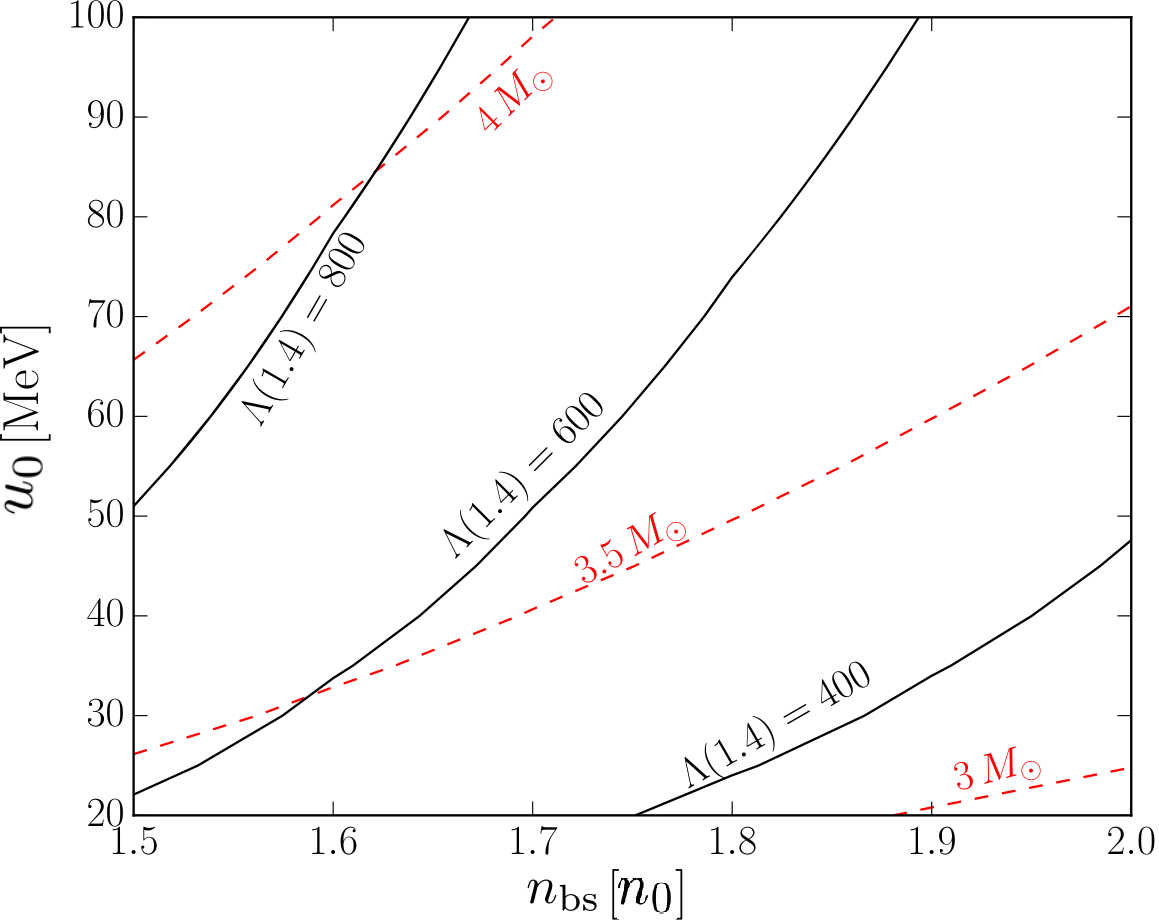}}
\caption{\label{fig:lambda} Constraints on the parameters of strangeon star models $u_0$ and $n_{\rm bs}$ (in unit of $n_0$). The tidal deformability of a 1.4 $M_\odot$ star $\Lambda (1.4)$ are plotted in solid lines, while that of the TOV maximum mass are shown in dashed lines. According to the constraint of GW 170817, any parameter choices below the top left solid curves is reasonable. Taken from Ref.~\citen{Lai2019_EPJA55-60}.}
\end{figure}

\begin{figure}[t]
\centerline{\includegraphics[width=0.7\linewidth]{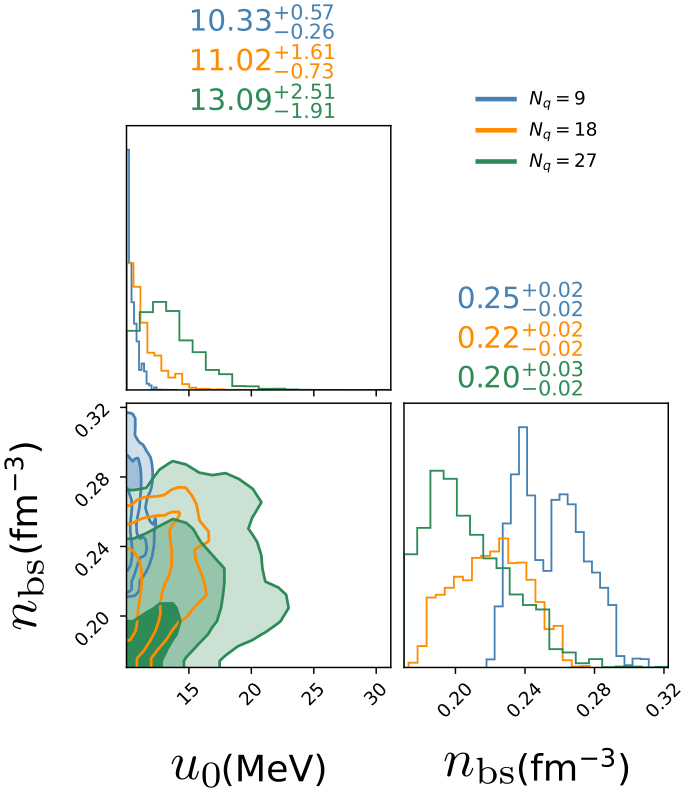}}
\caption{\label{fig:posterior} Posterior distribution of the model parameters under the constraints of PSR J0030+0451, PSR J0740+6620, and PSR J0437-4715 at different cases of fixed $N_q$. The contour levels in the corner plot correspond to the 68.3\%, 95.4\%, and 99.7\% confidence levels, going from dark to light. Taken from Ref.~\citen{Yuan2025_PRD111-063033}.}
\end{figure}

As indicated in Fig.~\ref{fig:lambda}, strangeon star models can satisfy both $M_{\rm TOV}>2.3 M_\odot$ and $\Lambda$ values smaller than the upper limit from GW170817 over a wide mass range~\cite{Lai2019_EPJA55-60}. In addition, as indicated in Fig.~\ref{fig:posterior}, Bayesian analysis was further carried out in the framework of Lennard-Jones model, where the potential depth $u_0$ and surface baryon density $n_{\rm bs}$ are constrained according to the latest mass-radius measurements of PSR J0030+0451, PSR J0740+6620, and PSR J0437-4715~\cite{Riley2019_ApJ887-L21, Riley2021_ApJ918-L27, Choudhury2024_ApJ971-L20}. It is found that the existing observational data indicates that the number of quarks inside a strangeon $N_q=18$ is more favorable~\cite{Yuan2025_PRD111-063033}.

\subsubsection{\label{subsubsec:merge}Merger: hydrodynamic simulations}

A full general-relativistic hydrodynamical evolution simulations for equal-mass binary strange star mergers with total masses of $\left(3.0 M_{\odot}, 3.05 M_{\odot}, 3.1 M_{\odot}\right)$ have been performed in Ref.~\citen{Zhou2022_PRD106-103030}. To verify whether prompt collapse occurs, the study tracked the evolution of the minimum lapse function $\alpha_{\rm min}$. For equal-mass binaries, the bounce of $\alpha_{\rm min}$ is associated with the bounce of  the remnant in the postmerger, which in turn drives a significant fraction of mass ejection during the merger. For prompt collapse, no such bounce occurs and the remnant straightforwardly collapses into a black hole, where the ejecta becomes negligible. Then one could identify the prompt collapse by searching for the least massive binary strange stars where no bounce occurs in $\alpha_{\rm min}$, e.g., as was done in Ref.~\citen{Zhou2022_PRD106-103030}.

\begin{figure}[t]
\centerline{\includegraphics[width=0.9\linewidth]{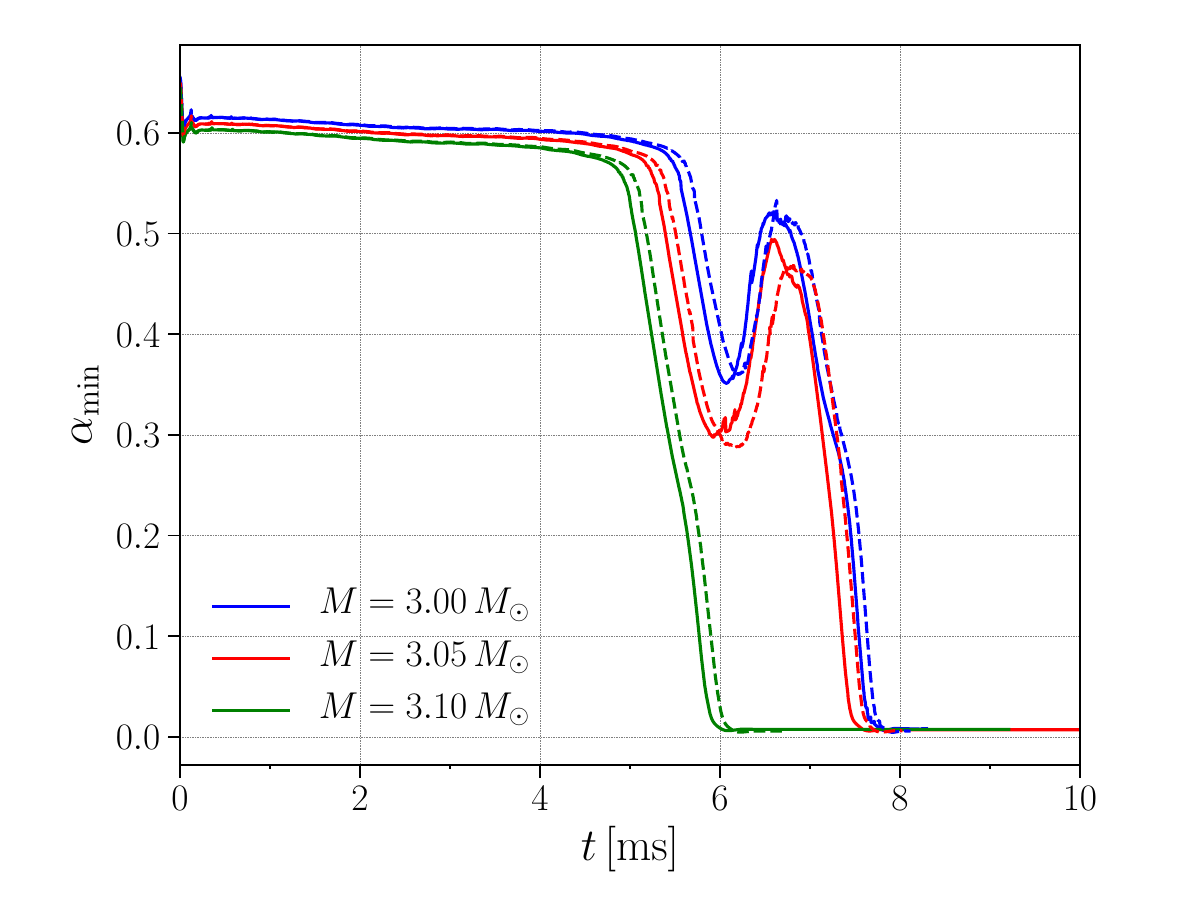}}
\caption{\label{fig:alpha_hr} The evolution of the minimum lapse function for three strange star mergers with total masses $\left(3.0 M_{\odot}, 3.05 M_{\odot}, 3.1 M_{\odot}\right)$. The results in the low and high resolution are shown in the dashed and solid curves, respectively. Taken from Ref.~\citen{Zhou2022_PRD106-103030}.}
\end{figure}

Figure~\ref{fig:alpha_hr} shows the evolution of $\alpha_{\min}$ during the mergers of binary strange stars with different total masses and resolutions. For binary strange star mergers with total masses of 3.0 $M_\odot$ and 3.05 $M_\odot$, $\alpha_{\min}$ exhibits a bounce before collapsing into a black hole. In contrast, for the case with a total mass of 3.1 $M_\odot$, the post-merger remnant collapses directly into a black hole without any bounce, constraining the threshold mass $M_{\rm{thres}}$ for this EOS model to lie in between 3.05 $M_\odot$ and 3.10 $M_\odot$, which is consistent across the two different resolutions. The lifetime of the post-merger remnant before collapsing into a black hole can be represented by the time difference between the first local minimum of $\alpha_{\min}$ and the moment when it approaches zero. According to Fig.~\ref{fig:alpha_hr}, the remnants of binary strange star mergers with total masses of 3.0 $M_\odot$ and 3.05 $M_\odot$ have similar lifetimes, approximately 1 ms, while the merger with a total mass of 3.10 $M_\odot$ corresponds to the prompt collapse case.

\begin{figure}[t]
\centerline{\includegraphics[width=0.9\linewidth]{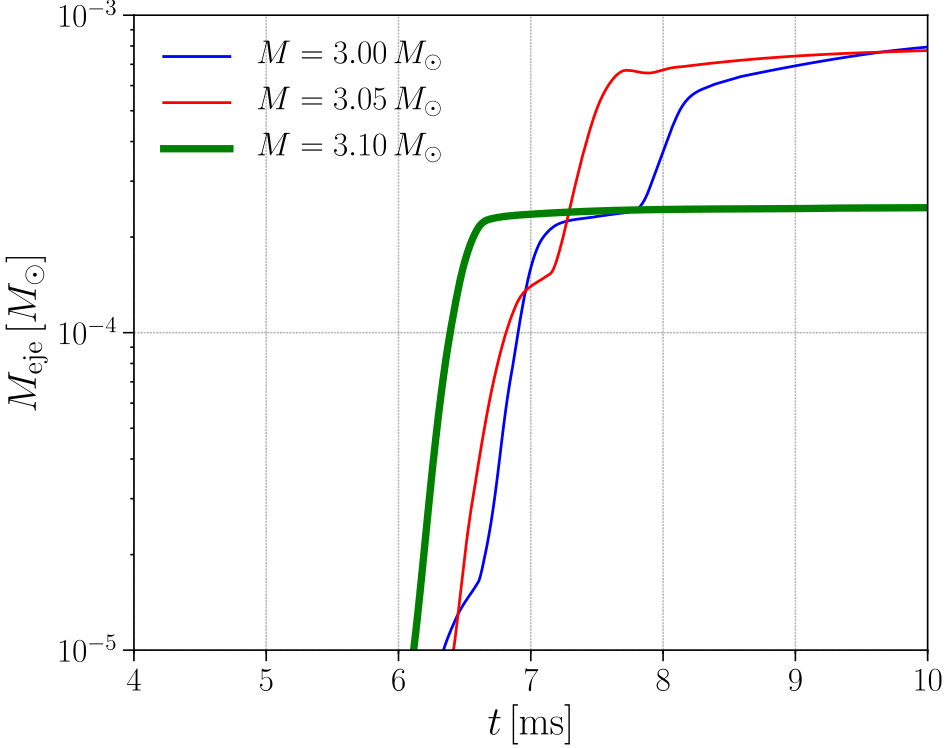}}
\caption{\label{fig:ejecta} Same as Fig.~\ref{fig:alpha_hr} but the amount of unbound material for three strange star mergers with total masses $\left(3.0 M_{\odot}, 3.05 M_{\odot}, 3.1 M_{\odot}\right)$, which are fixed in the high resolution runs. Taken from Ref.~\citen{Zhou2022_PRD106-103030}.}
\end{figure}

In addition to Fig.~\ref{fig:alpha_hr}, the time evolutions of the ejecta mass following the strange star mergers for the three cases are examined in Fig.~\ref{fig:ejecta}. Evidently, the ejecta mass is closely related to the postmerger outcome. For the binary strange star merger with a total mass of 3.10 $M_\odot$, ejecta appears earlier because the merger occurs earlier, but the final total ejecta mass is lower due to the absence of a bounce in the merger remnant. In contrast, when the remnant has a longer lifetime before collapsing into a black hole, the ejecta mass is larger due to the presence of a bounce. Meanwhile, for the binary strange star mergers with total masses of 3.00 $M_\odot$ and 3.05 $M_\odot$, the final ejecta masses are almost identical. These characteristics are expected to affect the electromagnetic radiation processes of binary strange star mergers, such as their associated short
gamma-ray burst and kilonova counterparts.

\subsubsection{Electromagnetic counterpart}

Besides measuring $\Lambda$ from mergers, we also study the kilonova emission and ejecta from binary strangeon star mergers~\cite{Lai2018_RAA18-024, Lai2021_RAA21-250}. The protons and neutrons evaporated from strangeon nuggets ejected during the merging process have different abundances at different temperatures, so they can produce two components, high opacity and low opacity ejecta. The obtained Bolometric light curve is consistent with the observational data~\cite{Lai2018_RAA18-024, Lai2021_RAA21-250}. Because strangeon stars have a large~$M_{\rm TOV}$, the post-merger remnant could be a long-lived supramassive strangeon star. Its spin-down energy release could power the X-ray plateau and internal plateau seen in some short GRBs~\cite{Yang2024_RAA24-035005}. Furthermore, in a strangeon star-black hole merger, pre-merger tidal cracking of the solid strangeon star could release enough energy from the starquake process, explaining the precursor of GRB 211211A~\cite{Zhou2024_RAA24-025019}.

\subsection{\label{subsec:DM}Strangeon Nuggets, Strangelets and Dark Matter}

\begin{figure}[t]
\centerline{\includegraphics[width=0.9\linewidth]{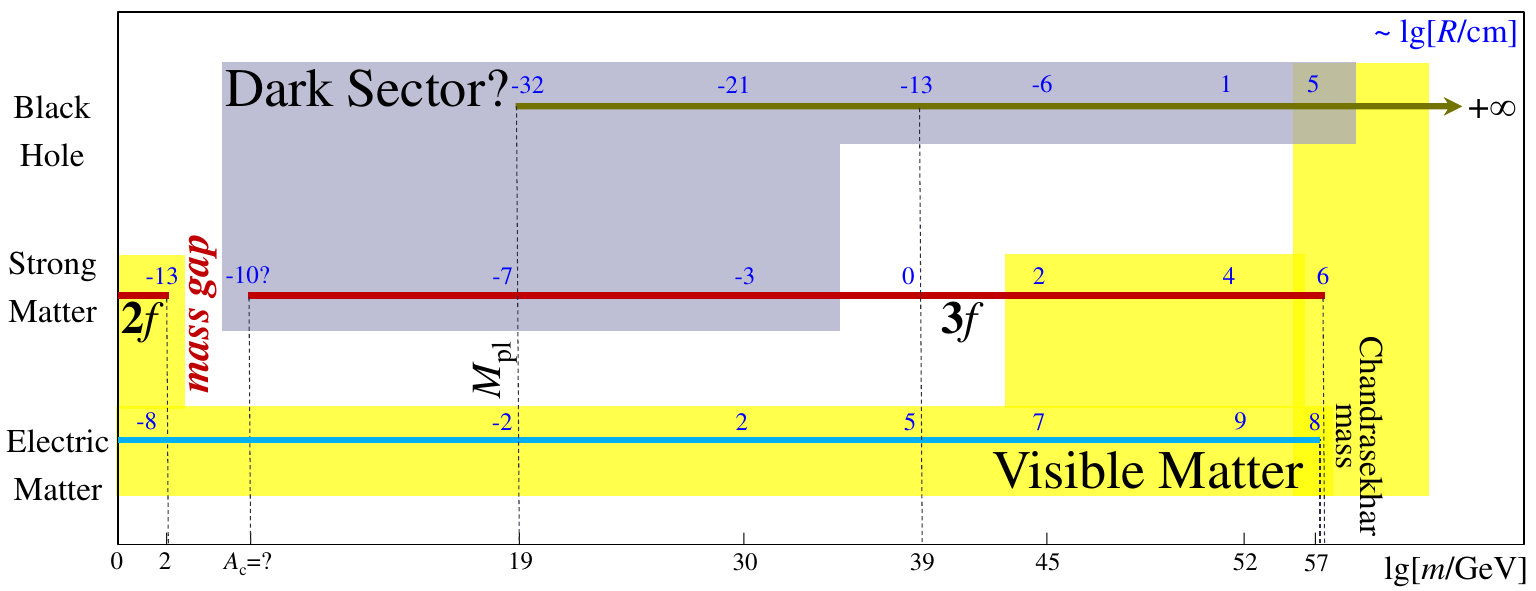}}
\caption{\label{fig:mass} Material world predicted by the standard model of particle physics. The mass spectra of baryonic matter, strangeon matter, and black holes are continuous, while their distributions are different. For strong matter, the nuclei contain two flavor quarks ({\bf 2}$f$), while the strangeon matter is of three flavor ({\bf 3}$f$). Taken from Ref.~\citen{Xia2024}. This plot for three types of matter offers insights into the material world structure in the Universe.}
\end{figure}

Strange matter may not only exist within compact stars but could also form nuggets populating the universe~\cite{Madsen2003_PRL90-121102}. It could also be a candidate for dark matter~\cite{Witten1984_PRD30-272}. Figure~\ref{fig:mass} illustrates the mass spectrum of compact objects under the strange matter hypothesis, where strangeon nuggets (masses between $10^{25}$ and $10^{35}$ g) or strangelets serve as dark matter candidates. They could help forming super massive black holes~\cite{Lai2010_JCAP2010-028}, which could be detected indirectly, e.g., using large neutrino telescopes~\cite{Qi2025_RAA25-095010}.

In particular, a strangeon nugget can be spontaneously magnetized~\cite{Lai2016_CPC40-095102}, which mainly originates from electron spin. As a fermionic system, electrons require an antisymmetric wave function. Consider two electrons without spin-orbit coupling: $\Psi(\Vec{r}_1,\Vec{r}_2,s_1,s_2) = \psi(\Vec{r}_1,\Vec{r}_2)\chi(s_1)\chi(s_2)$. For total spin $S = 1$, the spin part is symmetric, so the spatial wave function $\psi$ must be antisymmetric---vanishing at $\Vec{r}_1 = \Vec{r}_2$, reducing Coulomb repulsion and yielding lower interaction energy than the $S=0$ case. This favors parallel spins, leading to spontaneous magnetization and a large net magnetic moment $\mu$.

We now estimate the magnetic field. Before magnetization, electrons fill the Fermi sea with Fermi momentum $p_{\rm F} = \left(\dfrac{3n_e}{4\pi}\right)^{1/3} 2\pi \hbar$, where $n_e$ is the electron number density with $p_{\rm F} \gg m_ec$. After magnetization, the Fermi surfaces for spin-up and spin-down electrons shift to $p_{\rm F} \pm \Delta p$. This mismatch lowers electromagnetic energy but increases kinetic energy:

\begin{equation}\label{dEem}
\Delta E_{\rm em} = \xi \cdot \frac{1}{2} \frac{e^2}{\bar{d}_e}, \quad \text{with} \quad \bar{d}_e = \left(\dfrac{3}{4\pi}\right)^{1/3} \cdot \dfrac{2\pi \hbar}{p_{\rm F}},
\end{equation}

\begin{equation}\label{dEk}
\Delta E_{\rm k} \simeq 2\Delta p \cdot c,
\end{equation}
where $\xi > 1$ accounts for Coulomb interactions at small distances. At equilibrium ($\Delta E_{\rm em} = \Delta E_{\rm k}$):
\begin{equation}\label{dp}
\frac{\Delta p}{p_{\rm F}} = \xi \left(\frac{4\pi}{3}\right)^{1/3} \frac{\alpha_{\rm em}}{8\pi} = 0.064 \xi \alpha_{\rm em},
\end{equation}
with $\alpha_{\rm em} = e^2/(\hbar c) \simeq 1/137$. Since $\Delta p \ll p_{\rm F}$, only a small fraction of electrons align spins. The total magnetic moment is:
\begin{equation}\label{mu}
\mu\simeq \dfrac{4\pi p_{\rm F}^2 \cdot 2\Delta p}{4\pi p_{\rm F}^3/3}\cdot N_{\rm e}\mu_{\rm B},
\end{equation}
where $\mu_{\rm B} = e\hbar/(2m_e)$ and $N_e$ is the total
number of electrons in a strangeon nugget. The equatorial surface field is $B_0 = \frac{\xi\mu}{r_0^3}$, which is on the order of $8.4 \times 10^{11} \text{G}$, comparable to typical pulsar surface magnetic field strengths.

In such cases, when a strangeon nugget impacts ordinary matter, it deposits kinetic energy into the medium, which is converted to heat, ionizing the medium, and induces a shock wave as an acoustic signal. The interaction resembles the solar wind with Earth's magnetosphere, where the magnetopause radius is determined by balancing magnetic pressure and plasma pressure, i.e.,
\begin{equation}\label{rm}
r_{\rm m} = \left( \frac{B_0^2 r_0^6}{4\pi \rho_{\rm med} v^2} \right)^{1/6},
\end{equation}
with $\rho_{\rm med}$ being the density of medium and $v$ the relative velocity of the strangeon nugget. The magnetospheric cross-section is thus $\sigma_{\rm m} = \pi r_{\rm m}^2$, which represents the maximum effective cross-section of a strangeon nugget in ordinary matter and enables the detection of strangeon nuggets, e.g., via an acoustic array. Yes, let's try great efforts to search untraheavy stuff with underwater acoustic detectors in the coming years~\cite{Qi2025_RAA25-095010, Cleaver2025_PRD112-063060}.

\section{Conclusion and Outlook}

While nature exhibits broken symmetries, the strong interaction at the microscopic level (governing nucleons and nuclei) favors up and down quarks. The weak interaction rapidly converts strange quarks to up/down nucleons, making strangeness negligible in normal matter. However, under extreme conditions like compact star cores, to reduce the high Fermi energy of electrons, the system may restore flavor (u, d, s) symmetry. Considering the strong coupling, matter could exist as a Fermi liquid of quasi-particles (strange quark matter) or as a solid (strangeon matter). Strange matter, composed of up, down, and strange quarks, may exist as a stable state during the evolution of compact stars, observed as pulsars.

The internal structure of pulsars and the EOS of dense matter remain major challenges in nuclear and particle physics. Strange matter models bridge the gap between microscopic quark degrees of freedom and macroscopic astrophysical observations: achieving flavor symmetry restoration. These models are being tested against various pulsar observations and are expecting to be checked by more advanced facilities.

\section*{Acknowledgments}
This work was supported by the National SKA Program of China (Grant No. 2020SKA0120300 and No. 2020SKA0120100) and the National Natural Science Foundation of China (Grant No. 12275234).

%\bibliographystyle{ws-ijmpa}
%\bibliography{strange_quark}

\end{document}